\documentclass[aps, nofootinbib,  preprint, 12pt]{revtex4}
\usepackage{graphicx}

\newcommand{\be}{\begin{equation}}
\newcommand{\ee}{\end{equation}}
\newcommand{\bea}{\begin{eqnarray}}
\newcommand{\eea}{\end{eqnarray}}

\def\urltilde{\lower .7ex\hbox{\~{}}}

\def\lsim{\mathrel{\rlap{\lower4pt\hbox{\hskip1pt$\sim$}}
    \raise1pt\hbox{$<$}}}         
\def\gsim{\mathrel{\rlap{\lower4pt\hbox{\hskip1pt$\sim$}}
    \raise1pt\hbox{$>$}}}         

\usepackage{epsf,epsfig,subfigure,graphicx,amsmath,amssymb}
\usepackage{color}
\begin{document}
\draft

\preprint{IPMU09-0101}

\title{\Large Kaluza-Klein Dark Matter After Fermi}

\vspace{1.0cm}

\author{Chuan-Ren Chen${}^1$, Mihoko M. Nojiri${}^{123}$, Seong Chan Park${}^1$, Jing Shu${}^1$} 
\email{chuan-ren.chen@ipmu.jp,nojiri@post.kek.jp,seongchan.park@ipmu.jp,jing.shu@ipmu.jp}
\vspace{1cm}
\affiliation{\small ${}^1$Institute for the Physics and Mathematics of the Universe, The University of Tokyo, 
Chiba $277-8568$, Japan\\
${}^2$Theory Group, KEK, 1-1 Oho, Tsukuba, Ibaraki 305-0801, Japan\\
${}^3$The Graduate University for Advanced Studies (SOKENDAI),1-1 Oho, Tsukuba, Ibaraki 305-0801, Japan}

\vspace{1.0cm}

\begin{abstract}

Kaluza-Klein photon  in universal extra dimension models is one of the most attractive dark matter candidates as a weakly interacting massive particle.  Having a characteristic split spectrum in split universal extra dimension the relic density of Kaluza-Klein photon with $ 900 {\rm GeV}$ mass  is in good agreement with the observed dark matter amount in our Universe.  Interestingly Kaluza-Klein photon in the same mass range also provides excellent fits to the recently observed excesses in cosmic electron and positron fluxes. The amount of  gamma-ray contributions, mostly from tau decays, can be significant around 300 GeV, thus can be observed or constrained by the forthcoming Fermi-LAT diffuse gamma-ray data.

 \end{abstract}


\maketitle

\section{Introduction}

Dark matter problem is one of the most pressing problems in particle physics and astrophysics. Even though the presence of dark matter is firmly established by several independent experiments and observations, we still do not know what is the major component of dark matter. 

An attractive dark matter candidate is weakly interacting massive particle (WIMP) as it naturally reconciles with the standard thermal history of big bang cosmology. For being a proper WIMP candidate a particle is required to be a stable one (or its lifetime has to be much longer than the age of Universe) otherwise it could not contribute to the structure formation of Universe and its decay products should have been detected. Often a discrete symmetry is introduced to make a WIMP stable: R-parity in supersymmetric models, T-parity in Little Higgs models and Kaluza-Klein (KK) parity in extra dimension models. Once a discrete symmetry, e.g. $Z_2$, is conserved, the lightest $Z_2$ odd particle is automatically stable and can not decay to the standard model (SM) particles which are set to be even under the $Z_2$ transformation. 

Recently a particular interest has been attracted to Kaluza-Klein dark matter in univeral extra dimension model (UED) \cite{UED1, UED2, UED3, mUED DM1, mUED DM2, mUED DM3, mUED DM4, mUED DM5, Bergstrom:2004cy, Bergstrom:2004nr, Bringmann:2005pp, Barrau:2005au, Birkedal:2005ep} and its variety split-UED \cite{Park:2009cs, Chen:2009gz} since it has been shown that a TeV  Kaluza-Klein dark matter might be responsible for the excesses in positron fraction, flux of electron plus positron and gamma-ray in cosmic-ray data \cite{Hooper:2009fj, Bertone:2009cb, Okada:2009bz, Iltan:2009qj} from PAMELA \cite{Adriani:2008zr, Adriani:2008zq}, ATIC \cite{ATIC}, PPB/BETS \cite{PPB-BETS}, HESS \cite{Collaboration:2008aaa} and Fermi-LAT \cite{Collaboration:2009zk}. Indeed a TeV scale Kaluza-Klein photon, $B_1$, is naturally expected as a dark matter candidtate in UED  and it has advantages over a Majorana spinor dark matter (e.g. neutralino in MSSM) since no helicity suppression arises in annihilation into two light fermions. 

Split-UED has been suggested to address the problem of excessive hadron production  in conventional UED models by invoking a 5D bulk Dirac mass term for quarks which is generically allowed by symmetries of the theory in a way of keeping Kaluza-Klein parity intact \cite{Park:2009cs}. Production of hadron is suppressed by heavy Kaluza-Klein quarks so that the better fit to the PAMELA antiproton data is provided in split-UED.

In this paper we further consider Kaluza-Klein dark matter model with particular emphasis on new electron (here, electron means electron and positron since Fermi-LAT doesn't tell them apart.) and diffuse gamma-ray data of Fermi-LAT. Three issues are particularly interesting here. First, the smoother and further extending shape of $e^-e^+$ excess in high energy regime seems to suggest a rather heavier dark matter ($ \gsim$ 900 GeV) than what is suggested by ATIC/PPB-BETS ($\simeq 620$ GeV).  Second,  lack of peaky tip of flux also suggests the direct production of $e^-e^+$ channel  sub-leading.  Third, a sizable high energy  diffuse gamma-ray flux is expected in our model from inverse Compton scattering of electrons as well as tau decays. 

This paper is organized as follows. After a brief review of the model in Sec.~\ref{review}, we first consider the relic density of Kaluza-Klein photon in split-UED with splitting spectra of leptons and quarks in Sec.~\ref{sec:relic}. Concerning possible experimental constraints when we split fermions, we present bounds on 5D bulk masses for leptons in Sec.~\ref{split lepton}. Then we fit the positron fraction and the excess in $e^-e^+$ flux by introducing a larger Kaluza-Klein scale (Sec.~\ref{epm}). The diffuse gamma-ray excess in high energy regime ($\gsim 100$ GeV) is expected in split-UED, which could be seen by forthcoming Fermi-LAT diffuse gamma- ray detection (Sec.~\ref{Sec:gamma}). Then we conclude.

\section{Split-UED (review)}
\label{review}
Split-UED is an extra dimension model where a natural candidate of dark matter arises as a consequence of the symmetry of the theory, dubbed Kaluza-Klien parity. All the standard model fields are universally propagating through the bulk so that their Kaluza-Klein excited states provide phenomenological implications if the compactification scale is accessible by the experiments ($1/R\sim {\rm TeV}$). In the minimal setup, the gauge group is exactly the same as the one in the standard model, $G={\rm SU(3)_c\times SU(2)_W \times U(1)_Y}$, so KK gluons, weak gauge bosons and photons are involved. Quarks and leptons are also alleviated to higher dimensional ones and their KK spectra are doubled. One chirality of their zero modes are projected out by the orbifold condition, leaving the other one as the standard model fermions. The most prominent feature of split-UED is the presence of (double) kink masses for fermions while keeping the KK parity. 
Once 5D bulk mass is introduced, Kaluza-Klein modes get additional mass contributions and become heavier while zero mode remain massless because of the orbifold condition. Let us be more concrete. The five dimensional action of split-UED is the same as the minimal UED (MUED) except the 5D bulk mass terms:
\begin{eqnarray}
S_{\rm split-UED} = S_{\rm mUED} - \int d^4 x \int_{-L}^{L} \sqrt{g} \,\,m^{ij}_5(y)\bar{\Psi}_i \Psi_j
\end{eqnarray}
where $\Psi_i$ are the bulk Dirac spinors containing quarks and leptons in their zero modes and the kink mass, $m^{ij}_5(y) = \mu^{ij}_5 \theta(y)$, is introduced with a step function defined as $\theta(y>0)=1$ and $\theta(y<0)=-1$. One should notice that the zero mode wave functions, $\sim e^{\pm \int  m_5(y) dy}$,
are even functions under the inversion about the middle point of extra dimension, $y=0$, so that Kaluza-Klein parity is respected. Here $\pm$ signs are determined by chirality.

Klauza-Klein states of fermions, on the other hand,  get additional masses:
\begin{eqnarray}
m_n^2 = m_0^2  + k_n^2 + \mu_5^2
\end{eqnarray}
where the first term ($m_0^2$) comes from the ordinary SM Yukawa interaction, the second term ($k_n^2$) from the momentum of the extra dimension and the last term ($\mu_5^2$) from the 5D bulk mass. $k_n$ is determined by $\mu_5 = \pm k_n \cot k_n L$ for KK modes. Again $\pm$ sign depends on chirality. All the details of Kaluza-Klein decomposition are in the Appendix.

Having the generic idea of split-UED, we can control KK spectra by turning on some bulk mass parameters. In order to suppress the hadronic annihilation cross section and avoid the unnecessary flavor problems, we first choose our 5D bulk masses for quarks to be universal and larger than the typical KK scale, which is $\mu^{ij}_q = \mu_5 \,\delta_{ij}$. For the charged leptons, in order to control their annihilation cross section ratio among different flavors which are dominantly coming from the right-handed components, we choose separate 5D bulk masses $\mu_{e_R}$, $\mu_{\mu_R}$ and $\mu_{\tau_R}$ to achieve that. For the the left-handed leptons, we assume that their 5D bulk mass are generally small so that their couplings are still almost KK conserving. Therefore their contributions to the lepton flavor violation and four fermion operators are negligible.

\subsection{Relic abundance of Kaluza-Klein DM}
\label{sec:relic}

Relic abundance of Kaluza-Klein dark matter has been extensively studied (See \cite{mUED DM1} as a review on Kaluza-Klein dark matter and references therein). It is pointed out that due to coannihilation with the right-handed charged leptons (mainly $e_R^{(1)}$), the lightest KK particle (LKP) in  500 GeV $\sim$ 700 GeV mass range is needed to generate the right relic density in the original MUED~\cite{UED3, mUED DM1, mUED DM2, mUED DM3}. In split-UED, however, due to the presence of 5D bulk masses for the leptons (even when it is very small), the degeneracies between the LKP and the KK leptons are automatically removed. In this case, a LKP  mass between 900 GeV to 1 TeV will predict the right relic abundance \cite{mUED DM1, mUED DM2, mUED DM3} from 5 years WMAP data \cite{WMAP5yr} and is exactly the necessary value needed to fit the Fermi and Hess data.

It is possible to even get a heavier dark matter to account for the relic abundance in a slightly extended model. For instance, by turning on the brane kinetic term for the $SU(2)$ gauge bosons \cite{Flacke:2008ne}, it is possible that the $SU(2)$ KK gauge bosons have degenerate mass spectra with LKP. In this case, the coannihilation effects from the $SU(2)$ KK gauge bosons will increase the required masses to account for the right relic abundance up to 1.5 TeV \cite{mUED DM3}. 

One may wonder why the coannihilation with KK singlet leptons will decrease the dark matter mass for the right relic abundance while the coannihilation with $SU(2)$ KK gauge bosons will increase it. The reason is that the overall effective cross section, which roughly depends on $1/m_{DM}^2$, goes like 
\bea
\sigma_{eff} = \sum_{ij}^N \sigma_{ij} \frac{g_i g_j}{g_{eff}^2} \cdots \ ,
\eea
when coannihilation is included. 
When one particle species is removed from the degenerate spectra, not only we remove the relevant coannihilation channels which will decreases $\sigma_{eff}$, but we also decouple some particle species which will decrease $g_{eff}$ and increases $\sigma_{eff}$, so the overall effect is really a competition. The former wins in the coannihilation with the $SU(2)$ KK gauge bosons (one has to decrease the dark matter mass to compensate that so their relics density is fixed) while the latter wins in the coannihilation with the KK singlet leptons due to their small couplings.


\subsection{Splitting leptons}
\label{split lepton}
In this section, we consider that leptons also have their own 5D bulk masses in order to split their KK spectra and the relevant constrains. Because KK number is also violated in the lepton sector, the even KK gauge bosons will induce the four fermion operators between leptons and between leptons and quarks at tree level. The contact interactions between electrons and the SM fermions are parameterized by an effective Lagrangian, $\mathcal{L}_{eff}$ 
\begin{eqnarray}
\mathcal{L}_{eff} = \frac{g^2 \epsilon}{(1+\delta ) (\Lambda^{ef}_{ij \epsilon} )^2 } \sum_{i,j=L,R} \eta_{ij} \bar{e}_i \gamma_\mu e_i \bar{f}_j \gamma^\mu f_j \ ,
\end{eqnarray}
where $g^2$ is taken to be $4 \pi$ as the $\rho$ meson coupling. $\delta = 1 (0) $ for $f = e$ $(f \neq e)$, $\Lambda^{ef}_{ij \epsilon}$ is the scale of the contact interactions, $\epsilon = \pm 1$, $e_i$ and $f_{j} $ are left or right-handed spinors. 

Because all the even KK $Z$  bosons are almost completely in the weak gauge eigenstate (even KK $W^3$ boson), they only couple to the left-handed fermions, while the even KK photon couple to both the left-handed and right-handed fermions. So contact interactions involves right-handed fermions are less constrained since only even KK photon will contribute. Hence we only split the right handed electron KK spectrum to maximally suppress the electron annihilation branching ratio. 

In our setup, we only introduce the 5D bulk masses to localize of the SM fields at the center of the extra dimension \footnote{In our previous publications \cite{Park:2009cs, Chen:2009gz} we adopted the case of the SM fields localized toward boundaries in which there is a sign difference in our calculation.}. All the couplings between the SM fermons and even KK gauge bosons have the same signs therefore the interference is constructive except the case with right handed up quark ($u_R$) and the quark doublet ($Q_L$) due to the opposite sign of hypercharges of them to hypercharge of the right-handed electron ($e_R$).

\begin{figure}[t]
\includegraphics[width=0.43\textwidth, angle=0]{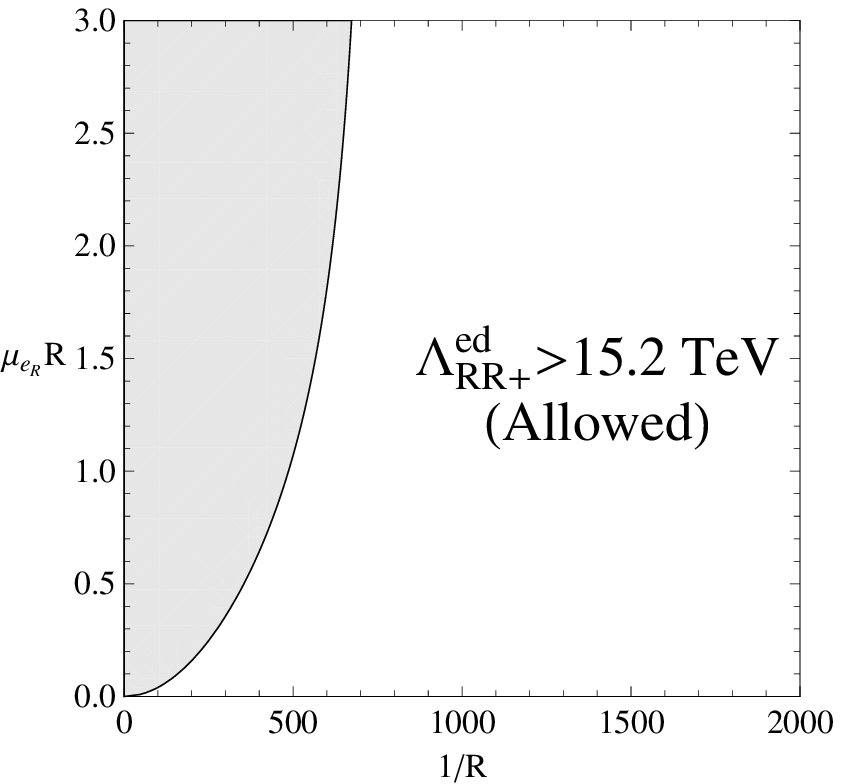}  
\includegraphics[width=0.4\textwidth, angle=0]{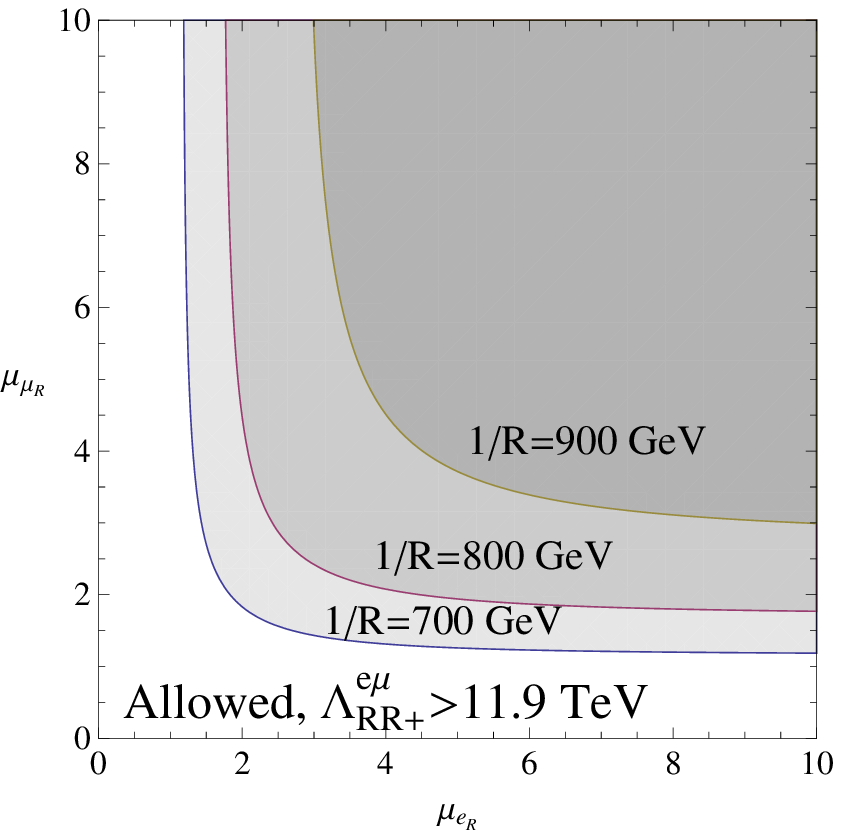}  
\caption{\label{fig:mubound} 
In the figure on the left we show the allowed space in ($\mu_{e_R}R, 1/R$) parameter space. The dark region is excluded by bounds for the contact $[e_R e_R d_R d_R]$ interaction. When $1/R>770$ GeV basically any value of $\mu_{e_R}$ is allowed. The figure on the right  shows the ($\mu_{e_R}, \mu_{\mu_R}$) parameter space with the KK scales $1/R=700, 800$ GeV and $900$ GeV where the dark region is excluded.
}
\end{figure}

We first consider the constrains for the 5D bulk mass of the right-handed electron. If the up or/and  the down quark has  5D bulk mass, then their  contact interactions with the electron will give the most stringent bound. 
From the global fit of the relevant data \cite{Barger:1997nf, Cheung:2001wx} when the $SU(2)_L$ gauge symmetry is assumed, by summing over the contributions from all even KK photons, we obtain the constrained parameter space for $\mu_{e_R}$ and $1/R$ in Fig. \ref{fig:mubound} from most stringent bound $\Lambda^{ed}_{RR+} = 15.2$ TeV \cite{Cheung:2001wx}  from the $e_R$-$d_R$ contact interaction 
\bea
\label{eq: eRdR}
\sum_{n} g^{e_R}_{2n00} (\mu_{e_R} R) g^{d_R}_{2n00} (\mu_{d_R} R) \Big [ \frac{g_1^2}{3 m^2_{B^{2n}}(R) } \Big ] \lesssim \frac{4 \pi }{(\Lambda^{ed}_{RR+} )^2}  \ . 
\eea
The terms of  $g^{e_R}_{2n00}$ and $g^{d_R}_{2n00}$ are the ratios of couplings between KK gauge bosons and SM fermions to the SM coupling for $e_R$ and $d_R$, respectively,
\begin{eqnarray}
g^{e_R}_{2n00} \equiv \frac{\int_{-L}^L dy f_{e_R^{(0)}} f_{e_R^{(0)}} f_{B^{(2n)}}  }{\int_{-L}^L dy f_{e_R^{(0)}} f_{e_R^{(0)}} f_{B^{(2n)}} },\,\, g^{d_R}_{2n00} \equiv \frac{\int_{-L}^L dy f_{d_R^{(2n)}} f_{d_R^{(0)}} f_{B^{(2n)}}  }{\int_{-L}^L dy f_{d_R^{(0)}} f_{d_R^{(0)}} f_{B^{(2n)}} }.
\end{eqnarray}

Here we use  $m_{B^{2n}}(R) \approx 2n/R$ and $g^{d}_{2n 00} \rightarrow \sqrt{2}$ by assuming all the KK quarks are decoupled. As we can see in Fig. \ref{fig:mubound}, the bound is quite loose and one can completely split the electron if the KK scale 
\bea
\frac{1}{R} > 770 ~ \mathrm{GeV} \ .
\eea

Clearly, this bound is satisfied for the dark matter mass used to fit the electron spectra in this paper, $m_{LKP} \geqslant 900$ GeV. 

The constraints on the 5D bulk mass of right-handed muon and tau are mainly coming from LEP \cite{Alcaraz:2006mx}. The bound for tau is relatively weak, and allows to introduce its arbitrary 5D bulk mass. The bound for 5D bulk mass of muon can be obtained from    
\bea
\label{eq: eRmuR}
\sum_{n} g^{e_R}_{2n00} (\mu_{e_R} R) g^{\mu_R}_{2n00} (\mu_{\mu_R} R) \Big [ \frac{g_1^2}{m^2_{B^{2n}}(R) } \Big ] \lesssim \frac{4 \pi }{(\Lambda^{e \mu}_{RR+} )^2}  \ ,
\eea
by taking $\Lambda^{e \mu}_{RR+} = 11.9$ TeV \footnote{Compare to $\Lambda^{e e}_{RR+} = 7.0$ TeV, $\Lambda^{e \tau}_{RR+} = 8.2$ TeV, the bound from muon is the most restrictive one \cite{Alcaraz:2006mx}.}. The allowed parameter space on $\mu_{\mu_R}$ vesus different $e_{\mu_R}$ is presented in Fig \ref{fig:mubound} when we fix the KK scale $1/R = 700, 800$ GeV and $900$ GeV, respectively. From the above discussions, we conclude that by introducing the 5D bulk masses for the right handed components of electrons and taus, we can adjust the their $B_1$ annihilation branching ratio arbitrarily, which can be used to fit the Fermi-LAT electron and gamma-ray data. 

\section{Cosmic-ray from KK dark matter}
The KK dark matter, $B_1$, mainly annihilates into fermion pairs among which charged lepton pair is dominant. Stable particles, like proton (p), electron ($e^-$) and photon ($\gamma$), and their antiparticles are generated from the casecade decay of hadrons and heavy charged leptons in the $B_1 B_1$ annihilation process and propagate to the Earth. In this section, we calculate the electron, positron and gamma-ray from the annihilation of KK dark matter and compare with the recent observations from PAMELA and Fermi-LAT collaborations. In the framework of split-UED, antiproton flux can be naturally suppressed and agrees well with the PAMELA data~\cite{Adriani:2008zq}, as seen in Ref.~\cite{Chen:2009gz}, therefore, we will not show the result of antiproton here.
%

\subsection{Electron and Positron}
\label{epm}
The calculation of flux of electrons and positrons is exactly the same as our previous work~\cite{Chen:2009gz}. Here we only give a brief review and some important equations in our analysis for completeness. The propagation of electrons and positrons ($e^{\pm}$) can be discribed by a diffusion equation as 
\begin{eqnarray}
K(E)\bigtriangledown^2 f_{e^\pm}(E,\vec{r}) + \frac{\partial}{\partial E}[b(E)f_{e^\pm}(E,\vec{r})] + Q(E,\vec{r}) = 0,
\end{eqnarray}
where $K(E)$ is the diffusion coeffieient, $f_{e^\pm}(E,\vec{r})$ is the density of $e^\pm$ per unit kinematic energy, $b(E)$ is the rate of energy loss and $Q(E,\vec{r})$ is the source of $e^\pm$. For annihilation of $B_1 B_1$, 
\begin{equation}
Q(E,\vec{r})=B_F \frac{1}{2}\left(\frac{\rho(\vec{r})}{m_{B_1}}\right)^2 \sum_i  \langle \sigma v \rangle_i \left(\frac{dN(E)_{e^\pm}}{dE}\right)_i,
\end{equation}
 where $dN_{e^{\pm}}/dE$ is the energy spectrum of $e^{\pm}$  
 obtained by using a Monte Carlo program, PYTHIA~\cite{Sjostrand:2006za}, and the index $i$ runs over all quark and charged lepton pairs, and $\rho(\vec{r})$ is the dark matter profile. In our numerical calculations, we adopt an overall boost factor $B_F$. There are known sources of boost factor: local clumps in dark matter profile \cite{clump1, clump2}, Sommerfeld enhancement effect by a  long range attractive force \cite{Sommerfeld} and  the Breit-Wigner type resonance effect \cite{Breit-Wigner}.  In split-UED,  without assuming a new attractive force or further tuning of mass spectra for large resonance effect, boost factor may mainly comes from local clumps.
 
The isothermal halo model \cite{Bergstrom:1997fj} is adopted as
\be
\rho_{\textrm{halo}} (\vec{r})=\frac{\rho_0}{1+(r/r_c)^2},
\ee
where $r=|\vec{r}|$  is the distance from our Galactic center, $r_c=3.5$ kpc and $\rho_0$ is the parameter that is adjusted to yield a dark matter local halo density of $0.3\, \rm{GeV}/cm^3$ \cite{Bergstrom:1997fj}
in our solar system.

Finally, the fluxes of electron and positron observed near the Earth is given as 
\begin{equation}
\Phi^{DM}_{e^{\pm}}(E)=\frac{c}{4\pi}f_{e^{\pm}}(E,r_{\odot}),\,
\label{eq:flux_e}
\end{equation}
where $c$ is the speed of light and $r_{\odot} = 8.5$ kpc is the distance from Solar system to the Galactic center.

\begin{figure}[]
\includegraphics[width=0.7 \textwidth, angle=0]{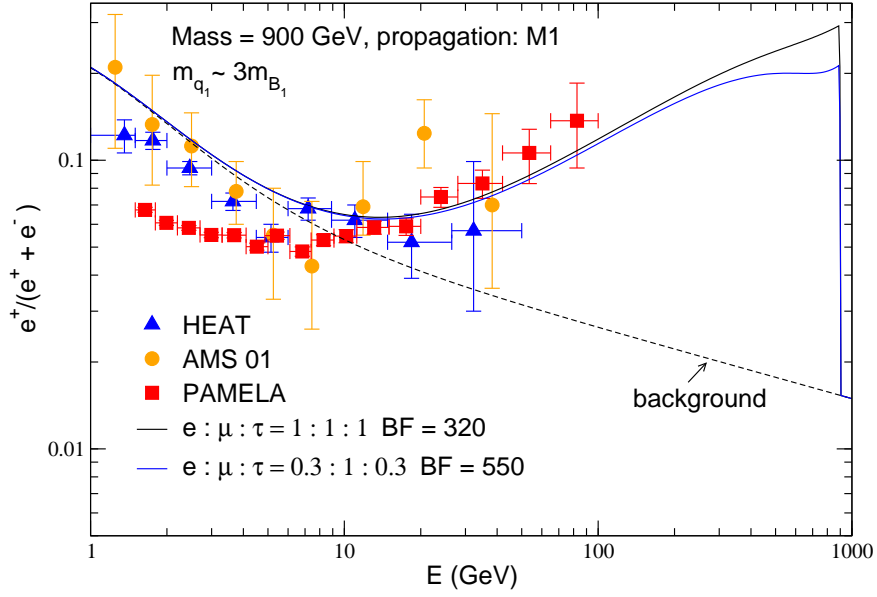}  
\caption{The positron fraction ($e^+/(e^+ + e^-)$) with $m_{B_1} = 900$ GeV and M1 propagation model, compared with PAMELA data. The mass of KK quark is taken to be about $3m_{B_1}$. The black curve is for $e:\mu:\tau=1:1:1$ case and the blue curve is for $0.3:1:0.3$ case. \label{pamela}.}
\end{figure}

\begin{figure}[]
\includegraphics[width=0.7 \textwidth, angle=0]{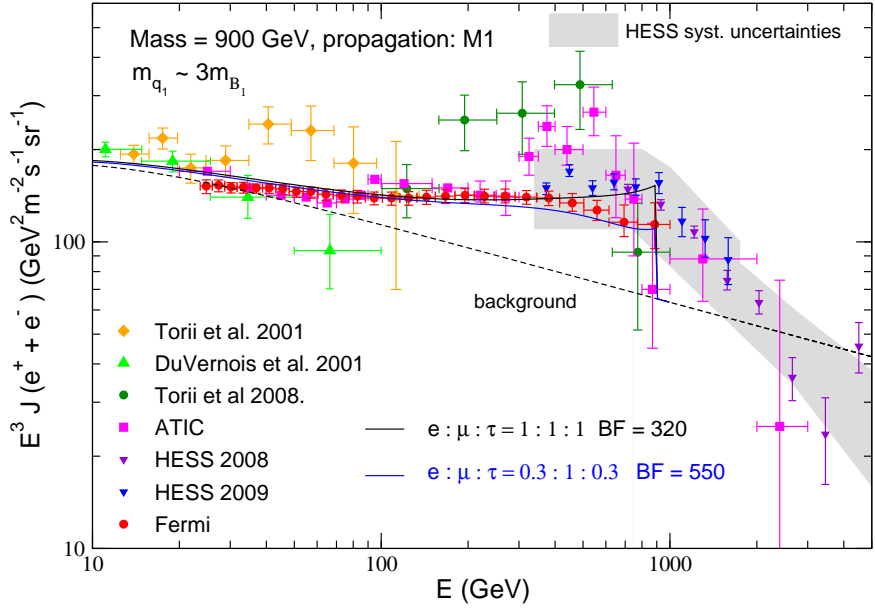}  
\caption{The total flux of positron and electron with $m_{B_1} = 900$ GeV and M1 propagation model, compared with Fermi-LAT and HESS data. The mass of KK quark is taken to be about $3m_{B_1}$. The black curve is for $e:\mu:\tau=1:1:1$ case and the blue curve is for $0.3:1:0.3$ case. \label{fermi}.}
\end{figure}

For the background of electron and positron, we adopt
approximations~\cite{Moskalenko:1997gh,Baltz:1998xv} of the fluxes of  primiary electron, secondary electron and secondary positron: 
\begin{eqnarray}
\Phi_{e^{-}}^{prim}(E) && =\frac{0.16E^{-1.1}}{1+11E^{0.9}+3.2E^{2.15}}\quad{\rm GeV}^{-1}{\rm cm}^{-2}{\rm sec}^{-1}{\rm sr}^{-1},\nonumber \\
\Phi_{e^{-}}^{sec}(E) && =\frac{0.7E^{0.7}}{1+110E^{1.5}+600E^{2.9}+580E^{4.2}}\quad{\rm GeV}^{-1}{\rm cm}^{-2}{\rm sec}^{-1}{\rm sr}^{-1},\nonumber \\
\Phi_{e^{+}}^{sec}  (E) &&=\frac{4.5E^{0.7}}{1+650E^{2.3}+1500E^{4.2}}\quad{\rm GeV}^{-1}{\rm cm}^{-2}{\rm sec}^{-1}{\rm sr}^{-1},\,
\label{eq:bg_e}
\end{eqnarray}
where $E$ is in units of GeV. We also use a free parameter $k$ multiplying the primiary electron backgound and we take $k=0.7$ in our numerical study. In Fig.~\ref{pamela} and Fig.~\ref{fermi}, we show the fracton of positron ($e^+/(e^+ + e^-)$) and the total flux of electron and positron ($e^+ + e^-$) from split-UED model with $m_{B_1} = 900$ GeV and compare with the experimental data, using so-called M1 propagation model~\cite{Delahaye:2007fr}. The mass of KK quark is split to be three times heavier than the $B_1$ dark matter through this paper below. The black solid line is for univeral KK leptons, namely all of the KK leptons have the same mass and are only slightly heavier than $B_1$. Therefor there exist equal amount of $e^\pm$, $\mu^\pm$ and $\tau^\pm$ in the final state.

 The case of $e:\mu:\tau=1:1:1$ fits the Fermi-LAT data acceptably well  with  $\chi^2_{dof} \lesssim 1$ \cite{Grasso:2009ma}.  Moreover, in the set-up of split-UED model, the spectrum of KK leptons can be also split  by turning on the bulk mass terms, and the fraction of different flavor of SM leptons in the final state will be different. Therefore, we can fit the Fermi LAT data even better if the rato is $e:\mu:\tau=0.3:1:0.3$ (in such case, the mass spectrum of KK lepton is $\tau_R^{(1)} \approx e_R^{(1)} > \mu_1 \approx B_1$) with the blue solid line in Fig.~\ref{fermi}. It can be seen that the peaky shape of the curve is much smoothen and fit the data very well. With such a parameter set, we can also fit the PAMELA data in Fig.~\ref{pamela}. In conclusion with a $900$ GeV KK dark matter, the anomalies of positron fraction and total flux of electron and positron can be well explained in the split-UED model.

\subsection{Gamma-ray}
\label{Sec:gamma}

Now, we consider the gamma-ray from the inner galactic plane (IGP, $0.25^\circ \leqslant |b| \leqslant 4.75^\circ$, $0.25^\circ \leqslant \ell \leqslant 29.75^\circ, 330.25^\circ \leqslant \ell \leqslant 360^\circ$) and the intermediate galactic latitudes ('10-20', $10^\circ \leqslant |b| \leqslant 20^\circ$, $0^\circ \leqslant \ell \leqslant 360^\circ$), both of which are observed by Fermi-LAT and the preliminary data have been presented in several conferences~\cite{plane, Porter:2009sg} , using the parameter sets that explain electron and positron data. The gamma-ray can be produced directly  through the decay of $\pi^0$ generated in the decay process of $\tau^\pm$ and the hadrons in the final state of $B_1 B_1$ annihilation. The latter is subdominant and can be neglected when the KK quark is much heavier than $B_1$, which is the case for satisfying the antiproton data of PAMELA. The flux per unit energy of these primary gamma-ray is given as
\begin{equation}
\left(\frac{dJ_{\gamma}}{dE}\right)_{primary}=\frac{1}{4\pi}\frac{1}{2m_{B_1}^2}\sum_i <\sigma v >_i \left(\frac{dN_{\gamma}}{dE}\right)_i \int_{los}\rho^2(\vec{\ell})d\vec{\ell},
\label{eq:gamma_halo}
\end{equation}
 where $i$ denotes the channel which produces photon in the final state and  $ \int_{los}\rho^2(\vec{\ell})d\vec{\ell}$
is the integration along the line of sight (los).

Other than  the primary gamma-ray mentioned above, the Inverse Compton Scattering (ICS) in the regions we are considering can also contribute to the energetic gamma-ray since there exists hard electrons and positrons in $B_1$ dark matter annihilation. And indeed, we will see later that the ICS can be comparable with the primary one. For ICS calculations, we follow the procedures in Ref.~\cite{Cirelli:2009vg} and a nice review of ICS can be found in Ref.~\cite{Blumenthal:1970gc}. The differential flux of the scattered photon with a energy $E$ is given as
\begin{equation}
\left(\frac{dJ_{\gamma}}{dE}\right)_{ICS}=\frac{1}{E}\frac{<\sigma v>}{4 \pi m_{B_1}^2} \int_{los}\rho(r)^2(\vec{\ell})d\vec{\ell} \int_{m_{e}}^{m_{B_1}} dE' \frac{{\cal P}(E,E',r)}{\dot{\cal E}(E',r)}Y(E'),
\label{eq:gamma_ics}
\end{equation}
where ${\cal P}(E,E',r) $ is the differential power emitted to a photon of energy $E$ by a positron or electron with energy $E'$, $\dot{\cal{E}}(E')$
is the total energy loss rate for a positron or an electron with energy $E'$ and $Y(E')$ is the total number of positron or electron with energy larger than $E'$. 
\begin{table}
\begin{tabular}{|c|c|c c c|}
\hline 
region&$\,\,\bar{J}\,\,$&$N_{SL}$ & $N_{IR}$ &  $N_{CMB}$\tabularnewline
\hline
\hline 
$0.25^\circ \leqslant |b| \leqslant 4.75^\circ$, &$10.0$&$1.7\times 10^{-11}$&$ 7.0\times 10^{-5}$&$1$\tabularnewline
$0.25^\circ \leqslant \ell \leqslant 29.75^\circ$,& & & &\tabularnewline
$ 330.25^\circ \leqslant \ell \leqslant 360^\circ$& & & &\tabularnewline
\hline
$10^\circ \leqslant |b| \leqslant 20^\circ$, &$ 2.3$&$8.9\times 10^{-13}$ &$1.3\times 10^{-5}$ &$1$\tabularnewline
$0^\circ \leqslant \ell \leqslant 360^\circ$ & & & & \tabularnewline
\hline
\end{tabular}
\caption{The geometrical factor $\bar{J}$ and parameters of modeling the Interstellar Radiation Field in Eq.(\ref{eq:n}). }
\label{tab:isrf}
\end{table}
The analytical expressions of $\cal{P}$ and $\dot{\cal{E}}$ are 
\bea
&&{\cal P}(E,E',r) =  \\ \nonumber
&&\frac{3\sigma_T}{4\gamma^2} E \int_{1/4\gamma^2}^{1} dq \left( 1-\frac{1}{4q \gamma^2 (1-\tilde{E})} \right) \frac{n(\epsilon,r)}{q}\left[ 2q \rm{ln} q +q+1-2q^2 +\frac{\tilde{E}^2}{2(1-\tilde{E})}(1-q) \right];\\ \nonumber
\\
&&\dot{{\cal E}}(E',r) = \\ \nonumber
&&3\sigma_T\int_0^\infty d\epsilon \epsilon\int_{1/4\gamma^2}^{1}dq n(\epsilon,r) \frac{(4\gamma^2-\alpha)q-1}{(1+\alpha q)^3} \left[ 2q \rm{ln} q +q+1-2q^2 +\frac{(\alpha q)^2}{2(1+\alpha q)}(1-q) \right],
\label{eq:PE}
\eea
where $\sigma_T = 8\pi r_e^2 /3=0.6625$ barn is the total Thomson cross section, $\tilde{E} = E/E'$ with $\gamma = E'/m_e$, $n(\epsilon,r)$  is the number density of background photon with energy $\epsilon$, $q=m_e\tilde{E}/(4 \epsilon \gamma (1-\tilde{E}))$ and $\alpha = 4\epsilon \gamma/m_e$. Approximately, for the region of Galaxy we are interested in,  the spatial dependence of Eq. (\ref{eq:gamma_ics}) can be integrated separately and  the simplified result reads 
\begin{equation}
\left(\frac{dJ_{\gamma}}{dE}\right)_{ICS}=\frac{1}{E}\frac{<\sigma v>}{4 \pi}r_{\odot}\frac{\rho^2(r_\odot)}{ m_{B_1}^2} \bar{J}\Delta \Omega \int_{m_{e}}^{m_{B_1}} dE' \frac{{\cal P}(E,E')}{\dot{\cal E}(E')}Y(E'),
\label{eq:ics_simp}
\end{equation}
where $\bar{J}$ is the geometrical factor, $\Delta \Omega$ is the solid angle of the observed region. The distribution of photon bath can be approximated by using the blackbody-like spectra~\cite{Cirelli:2009vg},
\begin{equation}
n_{a}(\epsilon) = \sum_i N_{a,i} \frac{\epsilon^2}{(\hbar c)^3}\frac{1}{e^{\epsilon/(kT_i)}-1},
\label{eq:n} 
\end{equation}
where the index $a$ represents the spacial region we are interested in and $i$ is for the different component of the photon bath, i.e. star light (SL), the infrared radiation (IR) from the galactic dusts which absorpt the star light and re-radiate out photons and photons of cosmic microwave background (CMB).The numerical values for IGP and '10-20' are listed in Table~\ref{tab:isrf} and the temparture $T_i$ for SL, IR and CMB are $0.3$ eV, $3.5$ meV and $2.725$ K, respectively~\cite{Cirelli:2009vg}.

\begin{figure}[t]
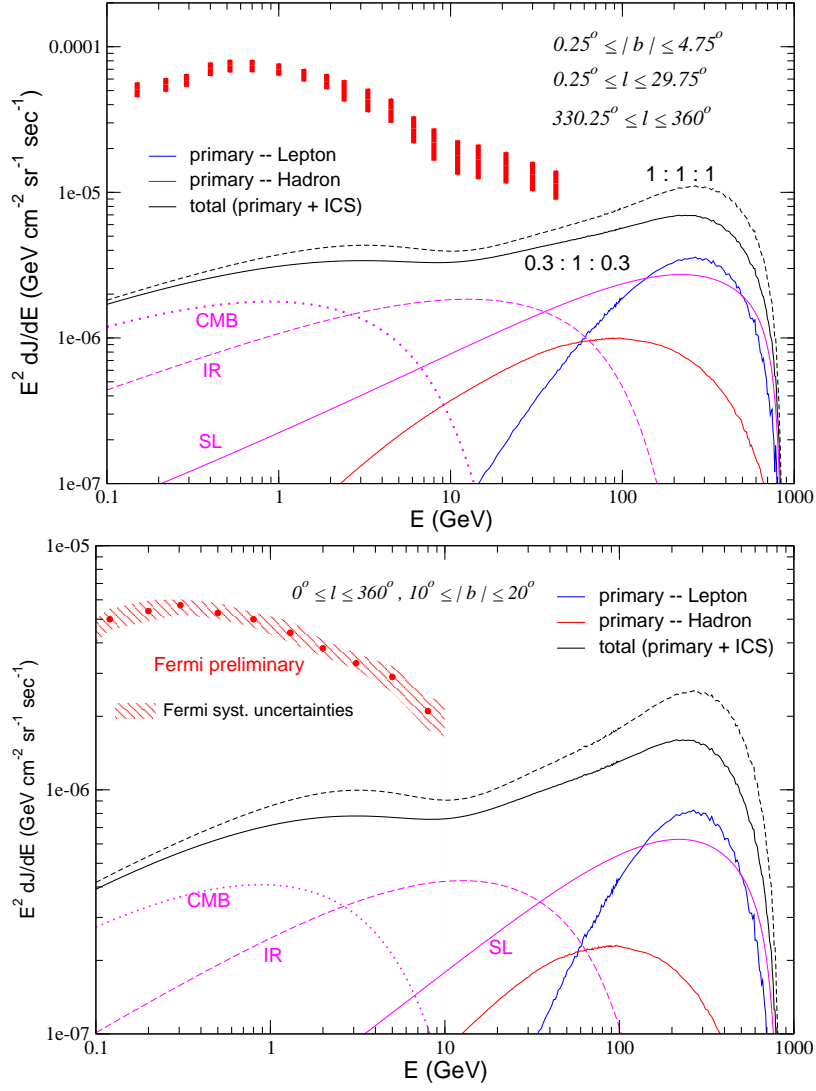

\includegraphics[width=0.65 \textwidth, angle=0]{gamma_gp.eps}\\  
\includegraphics[width=0.65 \textwidth, angle=0]{gamma_band.eps}  
\caption{The gamma-ray signal from $B_1$ dark matter annihilation for the regions of Inner Galactic plane (\textit{upper}) and intermediate galactic latitudes (\textit{lower}). The black solid and dashed curves are the sum of primary and ICS contributions when the final state charged lepton branching ratios are $(0.3:1:0.3)$ and $(1:1:1)$ respectively. The blue and red lines are the gamma-ray from $\tau^\pm$ and hadrons from $B_1$ dark matter annihilation, respectively, and the magenta lines are contributions of three components of ICS in the $(0.3:1:0.3)$ case.
\label{gamma}}
\end{figure}

In Fig.~\ref{gamma} we show the gamma-ray signal for Inner Galactic plane region and the region of intermediate galactic latitudes in the case that the KK leptons are not universal which has a better shape in flux of electron and positron shown in Fig.~\ref{fermi} and compare with the Fermi-LAT preliminary data. The KK quark mass spectra are taken to be $m_{q_1} \approx 3/R$ so that we are able to reach their KK quark production and decay signals at the LHC \cite{Chen:2009gz}. The black curve is the sum of primary and ICS contributions, the blue and red lines are the gamma-ray from $\tau^\pm$ and hadrons from $B_1$ dark matter annihilation, respectively, and the magenta lines are contributions of three components of ICS. In the region of low energy ($E\lesssim 10$ GeV), the dark matter signal is much smaller than the observed data, however, starting from few tens of GeV, the signal is about a factor of $2\sim3$ smaller. A characteristic of our model is that a bump at $E\approx 300$ GeV is predicted, due to the main contributions from primary $\tau^\pm$ and also the ICS from the star light in a subleading way, which can be checked soon if higher energy of gamma ray data is available. We also show the case of universa KK lepton in dashed line for reference. The behavior is similar to the non-universal case, but with more energetic gamma-ray due to more hard $e^\pm$ and more $\tau^\pm$. 
\section{Conclusion}

We consider the pair annihilation of the lightest Kaluza-Klein photons in split-UED as a primary source of recently observed cosmic ray positron and gamma in PAMELA and Fermi-LAT. Leptophilic property of dark matter suggested by the PAMELA antiproton data is naturally realized in split-UED. 

As the mass of dark matter particle around $900$ GeV and its primary annihilation channel being lepton pairs with $e:\mu:\tau=0.3:1:0.3$ (or $1:1:1$) we successfully fit the all the cosmic ray data. A particularly interesting prediction of our model is that the excess of cosmic gamma-ray flux, if observed by the forthcoming data of Fermi-LAT diffuse gamma-ray, peaks at $E\approx300$ GeV range. If there is no excess in the high energy region, then Fermi-LAT will put an upper bound on the tau fraction in our model.

Finally we point out another interesting prediction for the collider phenomenology. In the case of splitting right handed charged leptons (i.e. $0.3:1:0.3$ case) a large cross section of dilepton (in particular, $e_R e_R$ and $\tau_R \tau_R$) production is expected at the LHC through 2nd KK gauge boson exchanges. As these leptonic signals are rather clean we expect that the detection would be promising and we leave it for future study.


\section*{Acknowledgement}
We thank C. Csaki,  K. Hagiwara, T. Tait and F. Takahashi for useful discussions.
This work was supported by the World Premier International Research Center Initiative 
(WPI initiative) by MEXT, Japan. The work of J.S. was also supported by
the Grant-in-Aid for scientific research (Young Scientists (B)
21740169) from JSPS.

\section*{Appendix: KK decomposition in split-UED}
Let us consider a massive fermion on an orbifold $S^1/Z_2$ with two fixed points at $y=-L$ and $y=+L$:
\begin{equation} 
        \label{eq:BulkAction}
S = \int d^4 x \int_{-L}^{+L} dy 
\left[ i
 \bar{\Psi}\, \Gamma^M \partial_M \Psi
 - m_5(y) \bar{\Psi} \Psi  \right].
\end{equation} 
A $y$-dependent kink-mass is introduced for keeping KK-parity as
\begin{eqnarray}
m_5(y) = \mu ~\theta(y),
\end{eqnarray}
where $\theta(-L<y<0)\equiv -1$ and $\theta(0<y<L)\equiv +1$ and gamma matrices are  $\Gamma^M= (\gamma^\mu, i \gamma_5)$.  
Left (Right)-chiral fermion is defined as usual $\gamma_5 \Psi_{L/R}=\mp \Psi_{L/R}$ and a generic Dirac fermion is decomposed by $\Psi = \Psi_L +\Psi_R$. 

Varying the action with respect to $\bar{\Psi}_L$ and $\bar{\Psi}_R$ we obtain the standard bulk equation of motion which are given by
\begin{eqnarray}
i \gamma^\mu \partial_\mu \Psi_L - \gamma_5 \partial_5 \Psi_R - m_5 \Psi_R &=&0, \\
i \gamma^\mu \partial_\mu \Psi_R - \gamma_5 \partial_5 \Psi_L - m_5 \Psi_L &=&0
\end{eqnarray}
then using $\gamma_5 \Psi_{L/R}=\mp \Psi_{L/R}$ we finally get
\begin{eqnarray}
\left(\mp \partial_R -m_5\right)\Psi_{R/L} + i \gamma^\mu \partial_\mu \Psi_{L/R} =0.
\end{eqnarray}
%


Now we would like to discuss how to perform the Kaluza--Klein decomposition of these fields. 
In general, when the fermion belongs to a complex representation of the symmetry group, the KK modes can only acquire Dirac masses and the KK decomposition is of the form
\begin{eqnarray} 
        \label{eq:DiracKK}
\Psi_{L/R}=\sum_n \psi_{L/R}^n (x) f_{L/R}^n (y), 
\end{eqnarray} 
where $\psi_{L/R}^n$ are 4D  spinors which satisfy Dirac equations:
\begin{eqnarray}
i \gamma^\mu \partial_\mu \psi^n_{L/R}=m_n \psi^n_{R/L}.
\end{eqnarray}
Plugging this expansion into the bulk equations  we get the following set of coupled first order differential equations for the wave functions $f_{L/R}^n$: 
\begin{eqnarray} 
        \label{eq:1st}
\left(\mp \partial_5-m_5\right)f_{R/L}^n + m_n f_{L/R}^n=0.
\end{eqnarray} 

Applying $(\mp \partial_5 + m_5)$ on the first order equations we get a decoupled second order equations in the bulk:
\begin{eqnarray} 
0&=&(\mp \partial_5 +m_5)\left[(\mp \partial_5 -m_5)f_{R/L}^n +m_n f_{L/R}^n\right] \nonumber \\
&=&(\partial_5^2 -m_5^2 + m_n^2 \mp m_5')f_{R/L} \nonumber \\
&=&(\partial_5^2 +\Delta_n^2)f_{R/L},
 \label{eq:2nd}
\end{eqnarray} 
where $\Delta_n^2 \equiv m_n^2 -m_5^2 \mp m_5'$. 

\subsubsection{Zero mode solution: $m_0=0$}
For $n=0$, we can find a massless solution ($m_0=0$) rather easily. In the bulk ($y\neq 0$), the equations in \ref{eq:1st} are reduced to a simple first order equations:
\begin{eqnarray}
\left(\mp\partial_5-m_5\right)f_{R/L}^0=0
\end{eqnarray}
having simple solutions
\begin{eqnarray}
 f_{R/L}^0(y) \sim e^{\mp \int_{-L}^y m_5(y') dy'} \to
 f_{R/L}^0(y) = N_{R/L} e^{\mp \mu |y|}
 \end{eqnarray}
where the normalization factors are obtained by the normalization condition $\int_{-L}^L |f_{R/L}^0| ^2 =1$:
\begin{eqnarray}
N_{R/L}=\sqrt{\frac{\pm \mu}{1-e^{\mp 2\mu L}}}.
\end{eqnarray}

Depending on the sign of $\mu$ the shape of wave functions are determined. If $\mu>0$, e.g., $f_{R}^0$ is localized toward the middle point ($y=0$) and $f_L^0$ towards the end points ($y=\pm L$). 

\subsubsection{KK mode solution: Heavy modes ($m_n^2>\mu^2$)}
Depending on the sign of $\Delta_n^2$ the wave functions $f_{R/L}(y)$ will be either sines and cosines or sinhes and coshes. Here we first consider the case with $\Delta_n^2=k_n^2>0$.  In this case the KK modes are heavier than the bulk mass since $m_n^2 =\mu^2 + k_n^2>\mu^2$. We call them {\it heavy modes}.

 The wave equation for heavy modes looks simple as:
\begin{eqnarray}
(\partial_5^2 + k_n^2)f_{R/L}^n=0,
\end{eqnarray}
and their generic solutions are
\begin{eqnarray} 
        \label{eq:wv1}
f_{R/L}^n(y) = \alpha_{R/L}^n \cos k_n y + \beta_{R/L}^n \sin k_n y.        
\end{eqnarray} 
$\alpha$'s and $\beta$'s are related by Eq. \ref{eq:1st}:
\begin{eqnarray} 
&&\mp \alpha_{L/R}^n k_n - m_5 \beta_{L/R}^n+m_n \beta_{R/L}^n=0, \\
&&\pm \beta_{L/R}^n k_n -m_5 \alpha_{L/R}^n + m_n \alpha_{R/L}^n=0
\end{eqnarray} 
From the continuity condition at the middle point (${\rm lim}_{\epsilon \to 0}\left( f_{L/R}^n(-\epsilon) -f_{L/R}^n (+\epsilon)\right)=0$) we get a useful formula:
\begin{eqnarray}
\mu = \frac{\pm k_n (\beta_{L/R}^{n,>}-\beta_{L/R}^{n,<})}{2 \alpha_{L/R}^n}.
\label{continuity}
\end{eqnarray}
where we have used $\alpha^>=\alpha^< = \alpha$ from the continuity condition.

Now let us consider boundary conditions. We can have two independent choices of Dirichlet boundary conditions according to the $Z_2$ orbifold condition: $f_L (L) = 0 =f_L (-L)$ (DL) or $f_R(L)=0=f_R(-L)$ (DR).   
\begin{eqnarray}
0&=&f_{L/R}(L)= \alpha_{L/R}^{n,>} \cos k_n L + \beta_{L/R}^{n,>} \sin k_n L, \\
0&=&f_{L/R}(-L)= \alpha_{L/R}^{n,<} \cos k_n L - \beta_{L/R}^{n,<} \sin k_n L,
\end{eqnarray}
or
\begin{eqnarray}
\frac{\beta_{L/R}^{n,>}-\beta_{L/R}^{n,<}}{2 \alpha_{L/R}^n}= - \cot k_n L.
\label{BC:D}
\end{eqnarray} 
Now combining the continuity condition in Eq. \ref{continuity} and the Dirichlet boundary condition in Eq. \ref{BC:D} we get the master equation: $$\mu = \mp k_n \cot(k_n L)$$ for (DL/DR), respectively.  This equation determines the KK spectrum for heavy modes with any given values of $\mu$.

\end{document}